\documentclass{article} 
\usepackage{amssymb,latexsym,amsfonts}
\def\newpic#1{%
   \def\emline##1##2##3##4##5##6{%
      \put(##1,##2){\special{em:point #1##3}}%
      \put(##4,##5){\special{em:point #1##6}}%
      \special{em:line #1##3,#1##6}}}
\newpic{}
\newtheorem{teor}{Theorem}\newtheorem{theorem}[teor]{Theorem} 
 
\newtheorem{lemma}[teor]{Lemma} \newtheorem{proposition}[teor]{Proposition}

\newtheorem{cor}[teor]{Corollary} \newtheorem{piem}[teor]{Example}
\input cyracc.def
\font\tencyr=wncyr9
\def\cyr{\tencyr\cyracc}
\font\tencyi=wncyi9
\def\cyi{\tencyi\cyracc}

\begin{document}
\title
{The Lattice of  Machine Invariant \\ Sets and Subword 
Complexity}
\date{}
\maketitle
\title{}
\begin{abstract}
We investigate the lattice of machine invariant 
classes~\cite{buls}. This is an infinite completely distributive  lattice 
but it is not a Boolean lattice.
We show the subword complexity and the growth function create machine 
invariant classes.
\end{abstract}

\section{ Motivation} 
In different areas of mathematics, people consider a lot of hierarchies 
which are typically used to classify some objects according to their 
complexity. Here we formulate and discuss some hierarchies of machine 
invariant classes.

We are inspired by Yablonski's result~\cite{kudr}.
\begin{theorem} Every initial Mealy machine an ultimately periodic word 
transforms to an ultimately periodic word. Let $V=\langle Q,A,B,\circ ,\ast 
\rangle$, 
$q\in Q$, $|Q|=k$  and $x=uv^\omega$, $y=q*x=u'w^\omega$. Then  $|w|=\theta 
\tau$, where $\theta\backslash|v|$ and $\tau\in\{1,2,\ldots,k\}$. 
\end{theorem}

The invention and financial explotation of enciphering and deciphering 
machines is a lucrative branch of cryptography.  Until the 19th century  
they there mechanical; from the beginning of the 20th century automation 
made its appearance, around  the middle of the century came electronics 
and more recently microelectronic miniaturiziation. Today's microcomputers 
--- roughly the size, weight, and price of a pocket calculator --- have a 
performance  as good as the best enciphering machines from the Second Word 
War. That restores the earlier significance of good methods, which had been 
greatly reduced by the presence of `giant' computers in cryptanalysis 
centres~\cite{bau}. 

A {\em cryptosystem}~\cite{stin} is a five--tuple $\langle \mathcal{P, C, 
K, E, D}\rangle$, where the following conditions are satisfied:
\begin{itemize}
\item $\mathcal{P}$ is a finite set of possible plaintexts,
\item  $\mathcal{C}$ is a finite set of possible ciphertexts,  
\item $\mathcal{K}$, 
the keyspace, is a finite set of possible keys;
\item  for each $K\in\mathcal{K}$, there is an encription rule 
$e_K\in
\mathcal{E}$ and
\item  a corresponding decryption rule  $d_K\in\mathcal{D}$; 
\item each $e_K\,:\,\mathcal{P}\to\mathcal{C}$ and  
$d_K\,:\,\mathcal{C}\to\mathcal{P}$ are functions such that\\ $\forall 
x\in\mathcal{P}\; d_K(e_K(x))=x$.
\end{itemize}

This leads to the concept of a ciphering machine ~\cite{fom}.
A tuple\\ $\langle X, S, 
Y, K, z, f, g, h \rangle$ is called a {\em ciphering machine} if:
\begin{itemize}
\item $X$ --- a finite alphabet of possible plaintexts, 
\item $S$ --- a 
finite set of states of the ciphering machine, 
\item$Y$ --- a finite alphabet 
of possible ciphertexts, 
\item
$K$ --- a finite set of possible keys;
\item
$z\,:\,K \to S$, $f\,:\,S\times K\times X \to K$, $g\,:\,S\times K\times X 
\to S$, $h\,:\,S\times K\times X \to Y$ are functions.
\end{itemize}
Besides, it may be considered as  a special kind of a Mealy 
machine~\cite{fom}. Thus the Mealy machine appears in cryptography. This model, 
namely, Mealy machine, is being investigated intensively since the 
nineteen fifties (cf.~\cite{das, hart, pl, kurm, bar}).

Now more specifically. We shall describe one secret-key cryptosystem (Fig.~1).

\special{em:linewidth 0.4pt}
\unitlength 1mm
\linethickness{0.4pt}
\begin{picture}(115.00,60.00)(25,55)
\emline{40.00}{100.00}{1}{40.00}{70.00}{2}
\emline{40.00}{70.00}{3}{55.00}{70.00}{4}
\emline{55.00}{70.00}{5}{55.00}{100.00}{6}
\emline{55.00}{100.00}{7}{40.00}{100.00}{8}
\emline{80.00}{105.00}{9}{80.00}{85.00}{10}
\emline{80.00}{85.00}{11}{105.00}{85.00}{12}
\emline{105.00}{85.00}{13}{105.00}{105.00}{14}
\emline{105.00}{105.00}{15}{80.00}{105.00}{16}
\emline{80.00}{95.00}{17}{55.00}{95.00}{18}
\emline{120.00}{95.00}{19}{105.00}{95.00}{20}
\emline{55.00}{75.00}{21}{85.00}{75.00}{22}
\emline{40.00}{85.00}{23}{25.00}{85.00}{24}
\put(25.00,85.05){\vector(-1,0){0.2}}
\put(65.00,95.00){\vector(-1,0){0.2}}
\put(65.00,75.00){\vector(-1,0){0.2}}
\put(110.00,95.00){\vector(-1,0){0.2}}
\put(92.67,95.00){\makebox(0,0)[cc]{$\mathfrak{V}$}}
\put(47.00,85.00){\makebox(0,0)[cc]{$\mathfrak{S}$}}
\put(115.00,100.00){\makebox(0,0)[cc]{$x$}}
\put(70.00,100.00){\makebox(0,0)[cc]{$y$}}
\put(70.00,80.00){\makebox(0,0)[cc]{$p$}}
\put(32.00,90.00){\makebox(0,0)[cc]{$c$}}
\put(75.00,61.00){\makebox(0,0)[cc]{Figure 1.}}
\end{picture}

Let  $\mathfrak{S}$, $\mathfrak{V}$ be  devices represent respectively 
the
bitwise addition (modulo two) and a Mealy machine $V=\langle 
Q,A,\{0,1\},\circ ,\ast \rangle$. All users have identical devices. 

The 
plaintext and cryptotext spaces are both equal to $\{0,1\}^*$. First the 
users choose a key, consisting of $x\in A^\omega$. Every session of 
communication begins with the choice of a session key, namely, sender 
chooses $n\in \mathbb{N}$, $q\in Q$ and then sends those securely to 
receiver. Now sender computes $y=q*x[n,n+l]$, where $l+1$ is the length of 
plaintext $p$. The encription works in a bit-by-bit fashion, that is, 
$c_i=p_i+y_i(\mathrm{mod}2)$. 

When this is done, the security of the scheme of course depends in a crucial 
way on the quality of the $x\in A^\omega$ and the machine $V$. It is worth 
to mention at this stage of investigation this scheme serves only as extra 
(but important)
motivation for represented report, that is, why we examine 
infinite words with Mealy machines.

On the other hand if we restrict ourselves with finite words then we can 
state only: for every pair of words $u,v\in A^n$ there exists Mealy 
machine that transforms $u$ to $v$.  So we have a trivial partition of 
$A^*$.

\section{ Preliminaries} 
\normalsize In this section we present most of the notations and 
terminology used in this paper. Our terminology is more or less 
standard (cf.~\cite{luca}) so that a specialist reader may wish to 
consult this section only if need arise.

Let $A$ be a finite non-empty set and $A^*$ the free monoid generated by 
$A$. The set $A$ is also called an {\em alphabet}, its  elements  {\em 
letters} and those of $A^*$ {\em finite words}. The identity element of 
$A^*$ is called an {\em empty word} and denoted by $\lambda$. We set 
$A^+=A^*\backslash\{\lambda\}$.

A word $w\in A^+$ can be written uniquely as a sequence of letters as 
$w=w_1w_2\ldots w_l$, with $w_i\in A$, $1\le i\le l$, $l>0$. The integer 
$l$ is called the {\em length} of $w$ and denoted $|w|$. The length of 
$\lambda$ is 0. We set $w^0=\lambda\;\wedge\;\forall i\; 
w^{i+1}=w^iw\,.$ 

A word $w'\in A^*$ is called a {\em factor} (or {\em subword}) of 
$w\in A^*$ if there exist $u,v\in A^*$ such that $w=uw'v$. A word $u$ 
(respectively $v$) is called a {\em prefix} (respectively a {\em suffix}) 
of $w$.  A pair $(u,v)$ is 
called an {\em occurrence} of $w'$ in $w$.
A factor $w'$ is called {\em proper} if $w\ne w'$. We denote respectively by F$(w)$, 
Pref$(w)$ and Suff$(w)$ the sets of $w$ factors, prefixes and suffixes.

An (indexed) infinite word $x$ on the alphabet $A$ is any total map 
$x\,:\,\mathbb{N}\rightarrow A$. We  set for any $i\ge0$, $x_i=x(i)$
and write $$
x=(x_i)=x_0x_1\ldots x_n\ldots
$$
The set of all the infinite words over $A$ is denoted by $A^\omega$.

A word $w'\in A^*$ is a {\em factor} of $x\in A^\omega$ if there exist 
$u\in A^*$, $y\in A^\omega$ such that $x=uw'y$. A word $u$ 
(respectively $y$) is called a {\em prefix} (respectively a {\em suffix}) 
of $x$.  We denote respectively by F$(x)$, 
Pref$(x)$ and Suff$(x)$ the sets of $x$ factors, prefixes and suffixes. For 
any $0\le m\le n$, $x[m,n]$ denotes a factor $x_mx_{m+1}\ldots x_n$. An
indexed word 
$x[m,n]$ is called an {\em occurrence} of $w'$ in $x$ if $w'=x[m,n]$. The 
suffix $x_nx_{n+1}\ldots x_{n+i}\ldots$ is denoted by $x[n,\infty]$.

If $v\in A^+$ we denote by $v^\omega$ an infinite word
$$
v^\omega=vv\ldots v\ldots
$$ 
This word $v^\omega$ is called a {\em periodic word}. 
The {\em concatenation} of $u=u_1u_2\ldots u_k\in A^*$ and $x\in 
A^\omega$ is the infinite word 
$$
ux=u_1u_2\ldots u_kx_0x_1\ldots x_n\ldots
$$
A word $x$ is called {\em ultimately periodic} if there exist words $u\in A^*$, 
$v\in A^+$ such that
$x=uv^\omega$. In this case, $|u|$ and $|v|$ are called, respectively, an 
{\em anti-period} and a {\em period}.

A 3--sorted algebra $V=\langle Q,A,B, q_0, \circ ,\ast \rangle$ is called 
{\em an initial Mealy machine} if 
$Q,A,B$ are
finite, non"-empty sets, $q_0\in Q$; 
$\circ\,:\,Q\times A 
{\to}\,Q$ is
a total function and
$\ast \,:\,Q\times A {\to}B$ is
a total surjec\-tive function.
The mappings $\circ$
and $\ast$ may be extended to $Q\times A^*$  by defining
$$
\begin{array}{lr}
q\circ \lambda =q,\quad & q\circ (ua)=(q\circ u)\circ a \\
q\ast \lambda  =\lambda ,\quad & q\ast (ua)=(q\ast u)((q\circ u)\ast a)\,,
\end{array}
$$ 
for all $q\in Q$, $(u,a)\in A^*\times A$. Henceforth, we shall omit 
parantheses if there is no danger of confusion. So, for example, we will write
$q\circ u\ast a$ instead of $(q\circ u)\ast a.$ 

Let $(x,y)\in 
A^\omega\times B^\omega$. We write $y=q_0*x$ 
or $x\stackrel{V}{\rightharpoondown}y$ if $\forall n\; 
y[0,n]=q_0*x[0,n]$ and say machine $V$ {\em transforms} $x$ to $y$. 
We write $x{\rightharpoondown}y$ 
if there exists such $V$ that $x\stackrel{V}{\rightharpoondown}y$.

\section{ The Lattice of Machine 
 Invariant Sets} 

We say a word $x\in A_1^\omega$ is {\em apt} for  $V=\langle 
Q,A,B,q_0,\circ ,\ast \rangle$ if $A_1\subseteq A$. 
Let $\mathfrak{K}\ne\emptyset$ be any class of infinite words. The class 
$\mathfrak{K}$ is called {\em machine invariant} if every initial machine
transforms 
all apt words of $\mathfrak{K}$  to words of $\mathfrak{K}$.

{\em Remark}. If we like to operate with sets instead of classes 
then we may restrict ourselves with one fixed countable alphabet 
$\mathfrak{A}$=
$\{a_0, a_1, \ldots, a_n, \ldots\}$ and consider 
the set $\mathrm{Fin(}\mathfrak{A})$ of all non-empty finite subsets of 
$\mathfrak{A}$. Now the set $\mathfrak{K}$ may be chosen as the subset 
of $\mathfrak{F}=\{\,x\in A^\omega\,|\,A\in\mathrm{Fin}(\mathfrak{A})\,\}$.
Similarly, we may restrict ourselves with one fixed countable set 
$\mathfrak{Q}=\{q_1, q_2, \ldots, q_n, \ldots\}$ and consider 
only machines  from the set
$$\mathfrak{M}=\{\langle Q,A,B,q_0,\circ ,\ast \rangle\,|\,
Q\in\mathrm{Fin}(\mathfrak{Q})\,\wedge\,A,B\in\mathrm{Fin}(\mathfrak{A}) 
\}\,.$$
Thereby, the set 
$\emptyset\ne\mathfrak{K}\subseteq\mathfrak{F}$ is called {\em machine 
invariant} if every initial machine $V\in\mathfrak{M}$ transforms all apt 
words of 
$\mathfrak{K}$  to words of $\mathfrak{K}$.\smallskip

We follow the well established approach (cf.~\cite{dav}). 
For the reader's convenience, we briefly recall some basic definitions in 
the form appropriate for future use in the paper. 

Let $P$ be a set. An {\em order} on $P$ is a binary relation $\le$ on $P$ 
such that, for all $x, y, z\in P$ :  
\begin{itemize}
\item $x\le x$ --- reflexivity, 
\item $x\le y$ and $y\le x$ imply $x=y$ --- antisymetry, 
\item $x\le y$ and $y\le z$ imply $x\le z$ --- trnsitivity.
\end{itemize}

Let $S=\{s_i\,|\,i\in \mathcal{I}\}\subseteq P$ and 
$S^u=\{y\,|\,\forall s\in S\;s\le y\}$.  An element 
$x\in P$ is called a {\em join} of $S$ (we write $x=\cup S$ or 
$x=\cup_{i\in \mathcal{I}}s_i$) if $x\in S^u$ and $\forall s\in 
S^u\;x\le s$. We write $x\cup y$ instead of $\{x\}\cup\{y\}$.  Dually, 
let $S^l=\{y\,|\,\forall s\in S\;y\le s\}$ then an element $x\in P$ is 
called a {\em meet} of $S$ (we write $x=\cap S$ or $x=\cap_{i\in 
\mathcal{I}}s_i$) if $x\in S^l$ and $\forall s\in S^l\;s\le x$.
We write $x\cap y$ instead of $\{x\}\cap\{y\}$.

Let $P$ be a non-empty ordered set.
\begin{itemize}
\item An element $\bot\in P$ is called a {\em bottom}, if $\forall x\in P\; 
\bot\le x$. Dually, $\top\in P$ is called a {\em top}, if $\forall x\in 
P\; x\le\top$.
\item If $x\cup y$ and $x\cap y$ exist for all $x,y\in P$ then  $P$ 
is called a {\em lattice}.
 \item If $\cup S$ and $\cap S$ exist for all $S\subseteq P$ then 
 $P$ is called a {\em complete lattice}.
\end{itemize}

A complete lattice $L$ is said to be  {\em completely distributive},
if for any doubly indexed subset $\{x_{ij}\,|\,i\in\mathcal{I}, 
j\in\mathcal{J}\}$ of $L$ we have
$$
\bigcap\limits_{i\in\mathcal{I}}\, 
(\bigcup\limits_{j\in\mathcal{J}}x_{ij}\,)=
\bigcup\limits_{\alpha \,:\,\mathcal{I}\to\mathcal{J}} 
(\:\bigcap\limits_{i\in\mathcal{I}}x_{i\alpha(i)}\,)\,.
$$

Let $L$ be a lattice with $\bot$ and $\top$. For $x\in L$ we say $y\in L$ 
is a {\em complement} of $x$ if $x\cap y=\bot$ and $x\cup y=\top$. A  
lattice $L$ is called a 
 {\em Boolean} lattice if \begin{itemize}
\item for all $x,y,z\in L$ we have
$x\cap(y\cup z)=(x\cap y)\cup(x\cap z)$,
\item  $L$ has $\bot$ and $\top$, and
 each $x\in L$ has a complement $x'\in L$.
\end{itemize}

\begin{cor}\label{cor} \textup{\cite{buls}}
  Let $\mathfrak{L}$ be the set that
contains all machine invariant 
sets. Then $\langle \,
\mathfrak{L}, \cup, \cap \,\rangle$ is a completely  
distributive  lattice, 
where $\cup$, $\cap$ are respectively the set union and intersection. The 
bottom $\bot$   is the set 
of all ultimately periodic words, the top $\top=\mathfrak{F}$.
\end{cor}

An infinite word $x\in A^\omega$ is called a {\em recurrent} word if any
factor $w$ of $x$ has an infinite number of occurrences in $x$. Any word 
$x=uy$, where $u\in A^*$, $y\in A^\omega$ is called an {\em ultimately 
recurrent} word if $y$ is a recurrent word.

\begin{theorem}\label{th}\textup{\cite{buls}} Every initial Mealy machine 
an ultimately recurrent word transforms to an ultimately recurrent word. 
\end{theorem}
\begin{piem}\label{ex} \textup{
Let $x=(x_i)=1010^210^31\ldots0^n1\ldots$ then $x$ is not an ultimately 
recurrent word. Assume $\{a,b\}\cap\{0,1\}=\emptyset$. Let 
$y\in\{a,b\}^\omega$  be any  ultimately recurrent word but not an
ultimately periodic. Define $z', z''$ as follows:}
\end{piem}
$$
z'_i=\left\{\begin{array}{lr}
1, & \textup{if\/ } x_i=1\; \textup{and\/ } y_i=a,\\
y_i, & \textup{ otherwise;}
\end{array}\right.\qquad
z''_i=\left\{\begin{array}{lr}
1, & \textup{if\/ } x_i=1\; \textup{and\/ } y_i=b,\\
y_i, & \textup{ otherwise.}
\end{array}\right.
$$
The word $z'$ or $z''$ neither is ultimately periodic nor  ultimately 
recurrent.
Consider the Mealy machines $V_1$ and $V_2$ shown in Figure 2. Note
$z'\stackrel{V_1}{\rightharpoondown}y$ and 
$z''\stackrel{V_2}{\rightharpoondown}y$.\medskip

\special{em:linewidth 0.4pt}
\unitlength 1.00mm
\linethickness{0.4pt}
\begin{picture}(70.00,24.00)(4,130)
\put(8.00,145.00){\makebox(0,0)[cc]{$V_1$}}
\put(61.00,145.00){\makebox(0,0)[cc]{$V_2$}}
\put(16.00,145.00){\makebox(0,0)[cc]{$q_1$}}
\put(69.00,145.00){\makebox(0,0)[cc]{$q_1$}}
\put(27.90,141.00){\vector(-1,-1){0.20}}
\put(80.90,141.00){\vector(-1,-1){0.20}}
\put(33.50,145.00){\makebox(0,0)[lc]{1$ab$/$aab$}}
\put(86.50,145.00){\makebox(0,0)[lc]{1$ab$/$bab$}}
\put(47.50,134.00){\makebox(0,0)[lc]{Figure 2.}}
\put(20.00,145){\circle*{2}}
\put(73.00,145){\circle*{2}}
\put(25.00,145){\circle{10}}
\put(78.00,145){\circle{10}}
\end{picture}

So we have a method how to construct the infinite word that neither is 
ultimately periodic nor ultimately recurrent from an ultimately recurrent 
word if it is not ultimately periodic. We shall refer to this example in 
proof of such proposition.

\begin{proposition} $\mathfrak{L}$ is not a Boolean lattice.
\end{proposition}

{\em Proof}. Let $\mathfrak{K}=\{x\in\mathfrak{F}\,|\,x \; -
\mathrm{\;ultimately\; recurrent}\}$ then $\mathfrak{K}\in\mathfrak{L}$  by 
Theorem~\ref{th}. Suppose $\mathfrak{K'}\in\mathfrak{L}$ is a 
complement of $\mathfrak{K}$ then $\mathfrak{K}\cap\mathfrak{K'}=\bot$ and 
$\mathfrak{K}\cup\mathfrak{K'}=\mathfrak{F}$ by Corollary~\ref{cor}. Let 
$z\in\{z',z''\}$ such that $z\notin\mathfrak{K}$ (see Example~\ref{ex}) 
then $z\in\mathfrak{K'}$. Since $\mathfrak{K'}\in\mathfrak{L}$ and 
$z\rightharpoondown y$ (see Example~\ref{ex}) then $y\in\mathfrak{K'}$. 
Hence, $y\in\mathfrak{K}\cap\mathfrak{K'}=\bot$. Contradiction.

\section{ The Length }

Let $P$ be an ordered set. Then $P$ is called a {\em chain} or {\em 
totally ordered set}, if for all $x,y\in P$, either $x\le y$ or $y\le x$ 
(that is, if any two elements of $P$ are comparable).
If $C=\{x_0, x_1,\ldots,x_n\}$ is a finite chain in $P$ with 
$\mathrm{card}(C)=n+1$, then we say the {\em length} of $C$ is $n$. If 
$C$ is infinite chain in $P$, then we say the {\em length} of $C$ is 
$\mathrm{card}(C)$. The length of the longest chain in $P$ is called the 
{\em length} of $P$ and is denoted by $\ell(P)$. 

A machine $V=\langle Q_1\times Q_2, A_1, B_2,(q_1,q_2), \circ ,\ast \rangle$ 
is called a {\em series} of $V_1=\langle Q_1,A_1,B_1, q_1, 
\circ\hspace*{-0.7ex}\raisebox{0.5ex}{$'$} ,
\ast\hspace*{-0.7ex}\raisebox{0.5ex}{$'$} \rangle$ with
$V_2=\langle Q_2,B_1,B_2, q_2, \circ\hspace*{-0.9ex}\raisebox{0.5ex}{$''$} ,
\ast\hspace*{-0.9ex}\raisebox{0.5ex}{$''$} \rangle$ if 
\begin{eqnarray*}
(q',q'')\circ a&=&(q'\circ\hspace*{-1.2ex}\raisebox{0.5ex}{$'$}\:a,
q''\circ\hspace*{-1.4ex}\raisebox{0.5ex}{$''$}\:q'
\ast\hspace*{-1.2ex}\raisebox{0.5ex}{$'$}\:a),\\
(q',q'')\ast a&=&q''\ast\hspace*{-1.4ex}\raisebox{0.5ex}{$''$}\: q'
\ast\hspace*{-1.2ex}\raisebox{0.5ex}{$'$} \:a
\end{eqnarray*}
for all $(q',q'',a)\in Q_1\times Q_2\times A_1$.

\begin{lemma}\label{lemma1}   If $x\rightharpoondown y$ and 
$y\rightharpoondown z$ then $x\rightharpoondown z$.
\end{lemma}

{\em Proof}. Let $x\stackrel{V_1}{\rightharpoondown}y$ and  
$y\stackrel{V_2}{\rightharpoondown}z$. We can  choose machines  
$V_1=\linebreak\langle Q_1,A_1,B_1, q_1, 
\circ\hspace*{-0.7ex}\raisebox{0.5ex}{$'$} ,
\ast\hspace*{-0.7ex}\raisebox{0.5ex}{$'$} \rangle$ and
$V_2=\langle Q_2,A_2,B_2, q_2, \circ\hspace*{-0.9ex}\raisebox{0.5ex}{$''$} ,
\ast\hspace*{-0.9ex}\raisebox{0.5ex}{$''$} \rangle$ so that $B_1=A_2$. Then $V$ 
the series of $V_1$ with $V_2$ transforms $x$ to $z$.

\begin{cor} A set $V(x)=\{y\:|\:\exists V\in\mathfrak{M}\;
x\stackrel{V}{\rightharpoondown}y\}$, where $x\in A^\omega$ and 
\linebreak$A\in 
\mathrm{Fin}(\mathfrak{A})$, is machine invariant.
\end{cor}

{\em Proof}. Let $y\in V(x)$ and $y\rightharpoondown z$ then 
$x\rightharpoondown z$ by  Lemma~\ref{lemma1}. Therefore $z\in V(x)$.

\begin{cor} $\mathrm{card}(V(x))= \aleph_0$, where
$\aleph_0$ is the first infinite cardinality.
\end{cor}

{\em Proof}. Since $\mathrm{card}(\mathfrak{M})=\aleph_0$ then 
$\mathrm{card}(V(x))\le \aleph_0$. Note $\bot\subseteq V(x)$ by 
Corollary~\ref{cor}. Hence 
$\aleph_0=\mathrm{card}(\bot)\le\mathrm{card}(V(x))$. Therefore 
$\mathrm{card}(V(x))= \aleph_0$.\medskip

An order on $C$ is called a {\em well-ordering} on $C$ if $C$ is a chain and 
every subset $S\subseteq C$ has a {\em minimal element}, that is, 
$\exists\:\cap S\in S$. 

\begin{theorem}[Zermelo] For every non-empty set $C$ there exists a 
well-order\-ing on $C$.
\end{theorem}

\begin{proposition} There is a chain 
$\mathfrak{C}$ in $\mathfrak{L}$ such that $\mathrm{card}(\mathfrak{C})= 
\mathfrak{c}$, where $\mathfrak{c}=\mathrm{card}(\mathbb{R})$, $\mathbb{R}$  
---  the set of real numbers.
\end{proposition}

{\em Proof}.  The proof is an application of Zermelo's theorem. 

Let $A\in 
\mathrm{Fin}(\mathfrak{A})$ such that $\mathrm{card}(A)>1$ and $\preceq$ be 
any well-ordering on $A^\omega$, while $x \prec y$ means $x\preceq y$ and 
$x\neq y$. Then define ${\mathfrak K}(y)=
\bigcup_{x\preceq y}V(x)$ and a chain $\mathcal{I}=
\{y\:|\:\forall x\prec y\;\mathfrak{K}(x)\neq\mathfrak{K}(y)\}$ in 
$A^\omega$. Since $A^\omega$ is well-ordered there is the minimal element 
$x^{(1)}$ in $\mathcal{I}$.

Now suppose that $x^{(1)}\prec x^{(2)}\prec\ldots\prec x^{(k)}$ are the 
first $k$ elements of the chain $\mathcal{I}$. Since $\forall i\:
\mathrm{card}(V(x^{(i)}))=\aleph_0$ and ${\mathfrak K}(x^{(k)})=
\bigcup_{i=1}^kV(x^{(i)})$ then $\mathrm{card}({\mathfrak 
K}(x^{(k)}))=\aleph_0$. Since $\mathrm{card}(A^\omega)>\aleph_0$ then 
$\exists x\in A^\omega\:x\notin{\mathfrak K}(x^{(k)})$. Hence, the chain 
$\mathcal{I}$ has 
at least the $k+1$-st element $x^{(k+1)}$. Therefore, we can say proceeded 
by induction that $\mathrm{card}(\mathcal{I})\ge\aleph_0$.

Since 
$\bigcup_{x\in\mathcal{I}}V(x)\supseteq A^\omega$ it must follow that 
$\mathfrak{c}=
\mathrm{card}(A^\omega)\le\mathrm{card}(\bigcup_{x\in\mathcal{I}}V(x))\linebreak=
\mathrm{card}(\mathcal{I})\le\mathfrak{c}$. 
Let $\mathfrak{C}=\{\mathfrak{K}(x)\:|\:x\in\mathcal{I}\}$ then 
$\mathfrak{C}$ is a chain in $\mathfrak{L}$ and 
$\mathrm{card}(\mathfrak{C})=\mathrm{card}(\mathcal{I})=\mathfrak{c}$.

\begin{cor}\quad  The length $\ell(\mathfrak{L})=\mathfrak{c}$ .
\end{cor}

\begin{cor} \quad $\mathrm{card}(\mathfrak{L})\ge\mathfrak{c}$ .
\end{cor}

\section{ Subword Complexity}

Let $A$ be an alphabet then for each $n\ge 0$ we denote by $A^n$ the set 
of all words of length $n$. The function $f_x(n)=card(A^n \cap 
\mathrm{F}(x))$, where $x\in A^\omega$, is called the {\em subword 
complexity} of the word $x$ (cf.~\cite{ber}). The {\em growth} function of 
the word $x$ is defined as 
$g_x(n)=\sum_{i=0}^{n}f_x(i)\,.$

Let $f$, $g$ be total  functions. We write $g=O(f)$, if there exists 
such $c>0$ that 
$\forall n\in\mathbb{N}\;|g(n)|\le c\,|f(n)|\,.$ 
Let $\emptyset\ne\mathfrak{K}\subseteq\mathfrak{F}$. We say the 
{\em subword complexity} of the set $\mathfrak{K}$ is $f$ if 
$\forall x\in\mathfrak{K}\;f_x=O(f)\,.$
Similarly, we say the {\em growth} function of the set $\mathfrak{K}$ is 
$f$ if $\forall x\in\mathfrak{K}\;g_x=O(f)\,.$
\begin{lemma}\label{lemma} Let $V=\langle Q,A,B, q_0, \circ ,\ast\rangle$ 
be any Mealy machine. If $x\stackrel{V}{\rightharpoondown}y$
then $\forall n\; f_y(n)\le |Q|\, f_x(n)\,.$
\end{lemma}

{\em Proof}. Let $x\stackrel{V}{\rightharpoondown}y$ and  $u\in F(x)$ 
 then 
there exist  $q\in Q$ and $v\in F(y)$ such that $q*u=v$. Since $q\in Q$, 
it follows that machine $V$ can transform the word $u$  to $|Q|$ distinct 
words $v$ at the very most. 

Let $v\in F(y)$ and $|v|=n$ then there exsit $u\in F(x)$ and $q\in Q$ such 
that $q*u=v$.  Hence, $u$ is trasformed to $v$. Note $|u|=|v|$. Therefore, 
$f_y(n)\le |Q|\, f_x(n)$. 

\begin{proposition}\label{apgalv} Let 
$f\,:\,\mathbb{N}\to\mathbb{R}$ be any 
 total function. 
\begin{itemize}
\item[\textup{(i)}]  If 
$\mathfrak{K}_1=\{x\in\mathfrak{F}\,|\, f_x=O(f)\}$  then 
$\mathfrak{K}_1$
is  the machine invariant set. 
\item[\textup{(ii)}]  If
$\mathfrak{K}_2=\{x\in\mathfrak{F}\,|\, g_x=O(f)\}$  then 
$\mathfrak{K}_2$
is the  machine invariant set.
\end{itemize}
\end{proposition}

{\em Proof}. (i) Let $x\in\mathfrak{K}_1$  then $\forall 
n\in\mathbb{N}\;f_x(n)\le c\,|f(n)|$ for some $c>0$. Let
$x\stackrel{V}{\rightharpoondown}y$, where  $V=\langle Q,A,B, q_0, \circ 
,\ast\rangle$, then  by 
Lemma~\ref{lemma} $f_y(n)\le|Q|\,f_x(n)\le c\,|Q|\,|f(n)|$. Hence 
$f_y=O(f)$, that is, $y\in\mathfrak{K}_1$. 

(ii) Let $x\in\mathfrak{K}_2$  then $\forall 
n\in\mathbb{N}\;g_x(n)\le c\,|f(n)|$ for some $c>0$. Let
$x\stackrel{V}{\rightharpoondown}y$, where  $V=\langle Q,A,B, q_0, \circ 
,\ast\rangle$, then $g_y(n)=\sum_{i=0}^nf_y(i)\le\sum_{i=0}^n|Q|\,f_x(i)=
|Q|\sum_{i=0}^nf_x(i)=|Q|\,g_x(n)\le c\,|Q|\,|f(n)|$. Hence $g_y=O(f)$, 
that is, $y\in\mathfrak{K}_2$.\medskip

\section{Conclusion}

We say a word $x\in\mathfrak{F}$ is {\em more complicated as} 
$y\in\mathfrak{F}$
if $$
\forall\mathfrak{K}\in\mathfrak{L}\:
(x\in\mathfrak{K}\Rightarrow y\in\mathfrak{K})\;\&\; 
\exists\mathfrak{K}\in\mathfrak{L}\:
(x\notin\mathfrak{K}\:\&\:y\in\mathfrak{K})\:. $$
So the lattice $\mathfrak{L}$ gives  classification of infinite words 
that covers some aspects of complexity.
It sems natural if we choose more complicate words as ciphers. 
Proposition~\ref{apgalv} comes up to our expectations that the lattice 
$\mathfrak{L}$ would serve as a measure of words cryptographic quality.

It is worth to mention the idea that a lattice would serve as 
a measure of quality comes from fuzzy mathematics~\cite{gog}.

At this moment of course we have 
recognized a few elements of $\mathfrak{L}$. Therefore the problem,
what is the structure of lattice $\mathfrak{L}$,  remains.


\end{document}